\documentclass[toc]{PoS}

\usepackage{
amsmath,amssymb,
fontenc,times,mathptmx, graphicx}

\title{Subtleties of Non-Abelian Gauge Theories in Cold-Atomic Lattices}

\ShortTitle{Subtleties of NA Gauge Theories in CA Lattices}

\author{\speaker{Peter Orland}
\\
        Baruch College and the Graduate School and University Center, the City University of New York, New York, NY 10010, USA\\
        E-mail: \email{orland@nbi.dk}}

\abstract{I point out two of the subtleties referred to in the title. The first is
that gauge-invariant magnetic systems may realized under 
general circumstances, as suggested
by a simple theorem. The second subtlety is
that care is needed to identify the field theory simulated by a 
cold-atomic lattice gauge system. Though the
simplest such model confines in $2+1$ dimensions, it has non-relativistic ``gluon" excitations. Time-reversal invariance is
spontaneously broken in this system. The confinement 
mechanism
is related to an extra U($1$) gauge invariance.There is a model, suggested long ago by D. Rohlich 
and me, which is known 
to have relativistic spin waves. One of the outstanding theoretical problems 
is a better determination of the energy-momentum
relation of spin waves in different magnetic gauge systems.
}

\FullConference{The 31st International Symposium on Lattice Field Theory (Lattice 2013)\\
		29 July - 3 August, 2013,\\
		Mainz, Germany}

\begin{document}

\section{Introduction}

The rapid advance of ultracold-atom technology \cite{lewenstein} has made atomic lattice systems with
dynamical gauge fields a serious prospect \cite{reviews}, \cite{ZCR}, \cite{TCLO}. Such systems, first proposed
by Horn \cite{H} and examined by Banks and Zaks \cite{BZ} are Heisenberg-type magnets. They were generalized
and given the names ``gauge magnets" \cite{OR} and ``quantum-link models" \cite{CW}, \cite{BCW}. I use the first moniker 
here, but they both mean the same thing. Several groups of people ``discovered" gauge 
magnets (including D. Rohrlich and me), without realizing 
they were not the first. Only a finite number of representations of the gauge group exist. I think it is fair to include the remarkable
proposed realization of the Kogut-Susskind
Hamiltonian lattice gauge theory \cite{ZCR} among these systems, because (for practical reasons) the latter has a finite number of states per link.

I want to direct your attention to two subtle
aspects of the subject. First, gauge magnets 
may describe general models of hopping, tightly-bound
particles. There is an old example \cite{OH}, but this is a special case of a more general theorem.

The second aspect is that non-Abelian gauge magnets 
are not necessarily Yang-Mills theories. The simplest example 
of an ${\rm SU}(2)\times {\rm U}(1)$ gauge magnet has non-relativistic
``gluons", whose Lagrangian can be found \cite{OR}. With a link term, there are two phases
\cite{BZ}; simple arguments indicate that both phases confine 
(there may be a deconfined phase in $3+1$ dimensions). Confinement, however, can be understood in
terms of the U($1$) part of the gauge group - the quarks in this theory must also have Abelian charge. There
is a simple model with relativistic gluons; though its Yang-Mills action 
may contain topological terms. We
need a better understanding of how the energy of a gauge magnet spin wave depends on its 
momentum. The coherent-state methods devised thus far \cite{OR} cry out for improvement.

Nothing in the previous paragraph is controversial. Finding the field theory
describing a quantum spin system is an art, not a science. The accepted wisdom
was that all translation-invariant XXX spin-$s$ chains are gapless. That wisdom was wrong. Haldane \cite{Hal} argued
that the integer-spin case is the O($3$) sigma model with $\theta=0$, which is gapped (but {\em some} 
integrable cases of integer spin chains {\em are} gapless \cite{Fad}).

Spin-wave methods can't yield the phase diagram of
a quantum system, but are useful for understanding ultraviolet 
behavior. They help identifying the field theory, though not the vacuum.

\section{Non-Abelian Gauge Magnets}

The simplest SU($2$) gauge magnets have four states on each link of a spacial latttice $(x,j)$. Each
state corresponds to a component
of a Dirac spinor. Operators $\gamma^{\mu}(x)$satisfy
the anticommutation relations $[\gamma^{\mu}, \gamma^{\nu}]_{+}=2\delta^{\mu \nu}$, where $\mu,\nu=0,1,2,3$. Products
of these operators provide the closed Lie algebra SU($4$), which can also act on the one-link space
of states. The other generators of this SU($4$) are the $4\times4$ operators 
$[\gamma^{\mu},\gamma^{\nu}]_{+}=2\delta^{\mu\nu}$, 
$\gamma^{5}=-{\rm i}\gamma^{0}\gamma^{1}\gamma^{2}\gamma^{3}$,
$\rho^{\mu}=-{\rm i}\gamma^{5}\gamma^{\mu}$ and 
$\sigma^{\mu \nu}=-\frac{\rm i}{4}[\gamma^{\mu},\gamma^{\nu}]_{-}\,$, $\Sigma^{b\; \pm}
=\frac{1}{4}\epsilon^{bcf}\sigma^{cf} \pm \sigma^{0b}$, where $b,c,f=1,2,3$.

The lattice gauge fields are quantum connection or parallel-transport operators, defined as
\begin{eqnarray}
V_{j}(x)= U_{j}(x)+\alpha_{j}(x)
U^{5}_{j}(x)=[\gamma_{j}^{0}(x)\otimes {\rm 1}\!\!{\rm l}-{\rm i}{\vec \gamma}_{j}(x)\cdot \otimes {\vec \tau}]
+\alpha_{j}(x)[\rho_{j}^{0}(x)\otimes {\rm 1}\!\!{\rm l}-{\rm i}{\vec \rho}_{j}(x)\cdot \otimes {\vec \tau}]
, \nonumber
\end{eqnarray}
where $\alpha_{j}(x)$ is an arbitrary complex number. The Hamiltonian describes an
${\rm SU}(2)\times {\rm U}(1)$ gauge theory. It has the form
\begin{eqnarray}
H=J\sum_{x,\;j\neq k}{\rm Tr}\;V_{j}(x)V_{k}(x+{\hat j}a)V_{j}(x+{\hat k}a)^{\dagger}V_{k}(x)^{\dagger}
+ K\sum_{x,j}\gamma^{5}_{j}(x). \label{hamil}
\end{eqnarray}
The triplet of SU($2$) Gauss' law operators and the single U($1$) Gauss' law operators are
\begin{eqnarray}
G^{b}(x)=\sum_{j=1}^{d}\,[\,\Sigma^{b\;+}_{j}(x)-\Sigma_{j}^{b\;-}(x-{\hat j}a)\,], \;\;
G^{5}(x)=\sum_{j=1}^{d}\,[\,\gamma^{5}_{j}(x)-\gamma^{5}_{j}(x-{\hat j}a)\,], \label{GLOps}
\end{eqnarray}
respectively.
The Gauss' law operators (\ref{GLOps}) commute with the Hamiltonian. Therefore the 
state of the entire system 
$\Psi(t)$ satisfies both 
$G^{b}(x)\Psi(t)=S^{b}(x)\Psi(t)$
and
$G^{5}\Psi(t)=S^{5}(x)\Psi(t)$, 
where the charges $S^{b}(x)$
and $S^{5}(x)$ are determined by the initial state $\Psi(0)$. Notice that (\ref{hamil}) is 
nontrivial when $K=0$.

In Reference \cite{OH}, a particular representation of the operators on links used was
$\gamma^{0}=(T^{+}+T^{-})\otimes {\rm 1}\!\!{\rm l}$, ${\vec \gamma}={\rm i}(T^{+}-T^{-})\otimes 2{\vec S},$
where $\vec S$ is the spin of a particle which can fill one of two vacancies on a link, and $T^{\pm}$ moves the
particle between the two vacancies. The particles are SU($2$) rishons, described for SU($N$)
gauge magnets in \cite{BCW}. The motivation of introducing these particles in Reference \cite{OH} was to
show how gauge magnets could arise in the low-energy limit of hopping particles on a lattice. A pictorial representation is:

\begin{picture}(80,80)(0,-40)

\linethickness{1mm}

\put(20,-3){$x$}
\put(35,0){{\line(1,0){100}}}
\put(50,0){{\circle*{15}}}
\put(120,0){\circle{15}}
\put(140,-3){$x+{\hat j}a$}

\put(270,0){{\line(1,0){100}}}
\put(285,0){\circle{15}}
\put(355,0){{\circle*{15}}}

\linethickness{0.8mm}

\put(210, 35){$T^{+}$}
\put(190,20){\line(1,0){50}}
\put(240,20){\line(-1,1){10}}
\put(240,20){\line(-1,-1){10}}

\put(210, -45){$T^{-}$}
\put(190,-20){\line(1,0){50}}
\put(190,-20){\line(1,1){10}}
\put(190,-20){\line(1,-1){10}}

\end{picture}

\vspace{10pt}

\noindent
where the solid circle is the spinning particle and the empty circle is a vacancy. This pictorial representation shows that
the SU($2$) and U($1$) Gauss' law are restrictions on the total spin  and the particle number, respectively,
adjacent to a lattice vertex \cite{OH}, \cite{BCW}. The U($1$) Gauss' law condition, for the case of no
sources, means that the arrangement of particles around a site satisfies the ``ice" or ``six-vertex"
condition in the physical states spanning the Hilbert space. This means that adjacent to one site $x$,  only
the following configurations (labeled 1 to 6) appear:

\vspace{20pt}


\begin{picture}(80,40)(15,0)

\linethickness{1mm}

\put(0,20){1}
\put(-2,0){{\line(1,0){64}}}
\put(30,-32){{\line(0,1){64}}}
\put(10,0){{\circle*{15}}}
\put(30,20){{\circle*{15}}}
\put(50,0){\circle{15}}
\put(30,-20){\circle{15}}

\put(72,20){2}
\put(70,0){{\line(1,0){64}}}
\put(102,-32){{\line(0,1){64}}}
\put(82,0){{\circle*{15}}}
\put(102,20){\circle{15}}
\put(122,0){{\circle*{15}}}
\put(102,-20){\circle{15}}

\put(144,20){3}
\put(142,0){{\line(1,0){64}}}
\put(174,-32){{\line(0,1){64}}}
\put(154,0){{\circle*{15}}}
\put(174,20){\circle{15}}
\put(194,0){\circle{15}}
\put(174,-20){{\circle*{15}}}

\put(216,20){4}
\put(214,0){{\line(1,0){64}}}
\put(246,-32){{\line(0,1){64}}}
\put(226,0){\circle{15}}
\put(246,20){{\circle*{15}}}
\put(266,0){{\circle*{15}}}
\put(246,-20){\circle{15}}

\put(288,20){5}
\put(286,0){{\line(1,0){64}}}
\put(318,-32){{\line(0,1){64}}}
\put(298,0){\circle{15}}
\put(318,20){{\circle*{15}}}
\put(338,0){\circle{15}}
\put(318,-20){{\circle*{15}}}

\put(360,20){6}
\put(358,0){{\line(1,0){64}}}
\put(390,-32){{\line(0,1){64}}}
\put(370,0){\circle{15}}
\put(390,20){\circle{15}}
\put(415,0){{\circle*{15}}}
\put(390,-20){{\circle*{15}}}

\put(390,-20){\bf .}

\end{picture}

\vspace{60pt}

\noindent
If a color source is present at $x$, then the number of adjacent particles is one or three and the 
spin state of the source and the adjacent particles is a singlet.

\section{The hopping-parameter expansion}

The Hubbard model has a lattice
Hamiltonian with nearest-neighbor hopping spin-1/2 particles, with a local Coulomb interaction. For
half-filling, with a repulsive Coulomb term, perturbation theory in the hopping 
term yields an effective Heisenberg antiferromagnet. 

There is a similar mechanism to produce a gauge magnet from a 
lattice model of hopping particles, in the low-energy limit \cite{OH}. Consider 
a (not necessarily regular) lattice of sites, at which a particle may sit. Suppose the particle is
has a vector index, making a vector $N$-plet. The particle creation and annihilation operators at a site $j$
can be written as $c^{\dagger}_{j,\alpha}$ and $c_{j,\alpha}$, where $\alpha=1,\;\dots,\;N$. The
spin or isospin at a site $j$ is
$S_{j}^{a}=\sum_{\alpha \beta}c_{j,\alpha}^{\dagger}(S^{a})_{\alpha \beta}c_{j,\beta}$, where the
$N\times N$ matrices $(S^{a})_{\alpha \beta}$ are generators of the symmetry group. The lattice is subdivided 
into cells or ``bags" on the lattice, labeled by $F$. These cells are connected sets
of sites. The sets are disjoint, covering the entire lattice. Each cell is surrounded by a red boundary in the figure: 

\begin{figure}[ht] 
\centering
\includegraphics[width=5in]{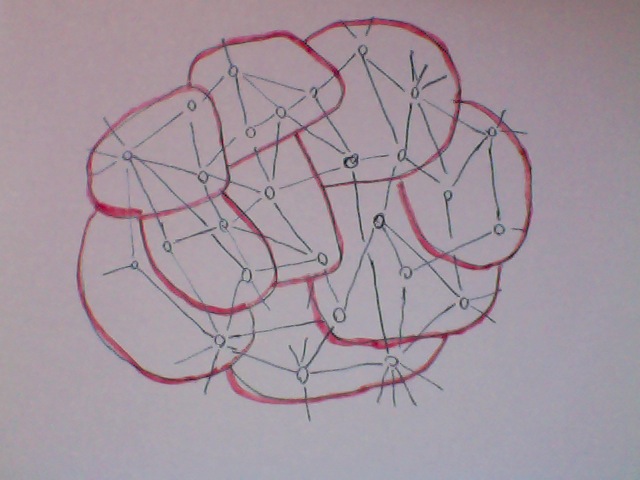} 
\nonumber
\label{masslessshell}
\end{figure}
The Hamiltonian has a nearest-neighbor hopping term, with hopping parameter $t$ and a term acting
on each cell $F$ of the lattice:
\begin{eqnarray}
H=-t\sum_{\langle i,j \rangle}\sum_{\alpha}c_{i, \alpha}^{\dagger}c_{j, \alpha}+U\sum_{F}V_{F}, \label{Cell-Site-H}
\end{eqnarray}
where $V_{F}$ must satisfy certain properties, namely: 
1. the eigenvector of 
$V_{F}$ with the smallest eigenvalue (or one-cell ground state) is
a singlet of the symmetry and
2. the eigenvectors of $V_{F}$ without the lowest eigenvalue are multiplets of the symmetry, with eigenvalues
at least of order $1$. Let's label the different multiplets by the letter $m=1,2,\dots$. There are a finite number of these, which depends
on the number of sites in one cell $F$. Since the multiplets are degenerate, the Hamiltonian commutes
with a projection operator $P^{m}_{F}$ onto the multiplet labeled by $m$.

\noindent
{\bf Theorem}: The effective Hamiltonian obtained in degenerate
perturbation theory in  
$t$, is a gauge magnet, where the Gauss law operator is 
\begin{eqnarray}
G_{F}^{a}=\sum_{j\in F}S_{j}^{a} -
\sum_{m}P^{m}_{F} \left(\sum_{j\in F}  S_{j}^{a}\right) \;P^{m}_{F}\;\sim \;({\vec D}\cdot {\vec E})^{a}-\rho^{a}.\label{GLT}
\end{eqnarray}
{\em Sketch of proof:} This operator annihilates the lowest energy states of the effective theory, provided $t\ll U$. The 
degenerate multiplets are color sources. Gauss' law $G^{a}_{F}\Psi=0$ is tautological (it is true by definition). The commutation
relations of the operators (\ref{GLT}) are precisely what generators of gauge transformations must
satisfy. Finally, taking the low-energy limit means that (\ref{GLT}) will commute with the effective
Hamiltonian. This effective Hamiltonian will have terms of
order  $t^{m}/U^{m-1}$ on polygons (plaquettes) with $m$ sides. A gauge-invariant
matter coupling may appear
at order $t^{2}/U$. {\em Note:} for the simplest gauge magnets, the role of the cell $F$ is played by the set of particle vacancies adjacent to
a lattice site.

The main implication of the theorem is that gauge invariance is nothing special. The theorem does not
guarantee that gauge invariance is truly dynamical, {\em i.e.}, the spin waves of the effective
theory have the quantum numbers of gluons\footnote{Indeed, Alessio Celi, Luca Tagliacozzo and I found
an example where gauge invariance is not dynamical (unpublished).}. Nonetheless, it does suggest that dynamical 
gauge invariance, {\em e.g.}, that of Reference \cite{OH},
can occur in a variety of contexts. 

I should mention that there is an alternative proposal for producing gauge magnets from Hubbard-type cold-atomic systems
\cite{LMT}.

\section{Confinement and U($1$) gauge symmetry} 

Suppose a single static quark is at $x$. Then Gauss' law implies that the number of particles on the link vacancies
adjacent to $x$ is one or three (so that total spin a singlet).

Let's consider the limit of (\ref{hamil}) as $K\rightarrow -\infty$. In this limit, only vertex 4 has finite energy. Notice that keeping only one such vertex breaks parity. This
means that a quark must produce a line of vertices going to the boundary (or an antiquark, if present) each
of which has energy of order $K$. Hence quarks are confined. In the illustration below, a quark-antiquark
pair forces a line of vertices other than vertex 4 to connect the sources. Thus confinement occurs with a string
tension of order $\vert K\vert/a$ \cite{TCLO}.

\begin{picture}(80,40)(0,0)

\linethickness{1mm}

\put(-20,0){{\line(1,0){380}}}
\put(10,-35){{\line(0,1){70}}}
\put(-10,0){\circle{15}}
\put(10,25){{\circle*{15}}}
\put(35,0){{\circle*{15}}}
\put(10,-25){\circle{15}}

\put(50,0){{\line(1,0){70}}}
\put(85,-35){{\line(0,1){70}}}
\put(60,0){\circle{15}}
\put(85,0){{\circle*{15}}}
\put(85,25){{\circle*{15}}}
\put(110,0){\circle{15}}
\put(85,-25){\circle{15}}

\put(120,0){{\line(1,0){70}}}
\put(160,-35){{\line(0,1){70}}}
\put(135,0){{\circle*{15}}}
\put(160,25){{\circle*{15}}}
\put(185,0){\circle{15}}
\put(160,-25){\circle{15}}

\put(195,0){{\line(1,0){70}}}
\put(230,-35){{\line(0,1){70}}}
\put(205,0){{\circle*{15}}}
\put(230,25){{\circle*{15}}}
\put(255,0){\circle{15}}
\put(230,-25){\circle{15}}

\put(270,0){{\line(1,0){70}}}
\put(300,-35){{\line(0,1){70}}}
\put(275,0){{\circle*{15}}}
\put(300,25){{\circle*{15}}}
\put(325,0){{\circle*{15}}}
\put(300,-25){\circle{15}}
\put(300,0){{\circle*{15}}}

\put(340,0){{\line(1,0){70}}}
\put(375,-35){{\line(0,1){70}}}
\put(350,0){\circle{15}}
\put(375,25){{\circle*{15}}}
\put(400,0){{\circle*{15}}}
\put(375,-25){\circle{15}}

\end{picture}

\vspace{35pt}

\noindent
Similarly,
in the limit that $K\rightarrow +\infty$, vertex 3 is dominant and a similar mechanism produces confinement.

This confinement mechanism is inherently Abelian. The argument works in U($1$) theories with electric charges
but no quarks. As $\vert K/J\vert$ decreases, there can be a phase transition to a phase in which one type of
vertex is no longer longer frozen into the system. It may be that this is the transition has been already found \cite{BZ}.
Is the second phase deconfined phase? I think it is clear that the answer is no. The reason is that discrete rotation
invariance is broken, in the same way that this happens in quantum-dimer models \cite{RK}, which
also have a U($1$) gauge invariance. There can be special choices of $K/J$
for which the correlation length becomes infinity where deconfinement occurs, but these are not generic. 

\section{Spin-wave frequency and wavelength}

The simplest model of the type (\ref{hamil}) is with $\alpha_{j}(x)=0$ everywhere:
\begin{eqnarray}
H=J\sum_{x,\;j\neq k}{\rm Tr}\;U_{j}(x)U_{k}(x+{\hat j}a)U_{j}(x+{\hat k}a)^{\dagger}U_{k}(x)^{\dagger}
+K\sum_{x,j}\gamma^{5}_{j}(x). \label{hamil1}
\end{eqnarray}
A way to study spin waves is to find the Heisenberg equation of motion ${\rm i}\partial_{t}B=[H,B]$ of a local
operator $B$, defined
on a single link. If $\gamma^{\mu}$ and $\rho^{\mu}$ are replaced, in the definition of $U_{j}$ and $U^{5}_{j}(x)$, by the classical variables $m^{\mu}$ and
$n^{\mu}$ respectively, with $n\cdot n=m\cdot m=1$, $m\cdot n=0$, these equations for $m^{\mu}$ and
$n^{\mu}$ follow from the classical action \cite{OR}, \cite{4DD}:
\begin{eqnarray}
S&=&\sum_{x,j} \; s\int dt \int_{0}^{\infty} du\;\; \epsilon_{\alpha \beta \mu \nu}\;n_{j}^{\alpha}m_{j}^{\beta}
\left( \frac{\partial n_{j}^{\mu}}{\partial t}\frac{\partial n_{j}^{\nu}}{\partial u}+
\frac{\partial m_{j}^{\mu}}{\partial t}\frac{\partial m_{j}^{\nu}}{\partial u}
\right)
\nonumber \\
&-&J\sum_{x,\;j\neq k}{\rm Tr}\;U_{j}(x,t)U_{k}(x+{\hat j}a,t)U_{j}(x+{\hat k}a,t)^{\dagger}U_{k}(x,t)^{\dagger}, \label{class}
\end{eqnarray} 
with spin $s=1/2$. The first term in (\ref{class}) is the Wess-Zumino action for relativistic spin \cite{4DD}. If 
the equations of motion are linearized, the spin wave frequency $E$ in terms of
its wave number $p$ is $\vert E\vert=4Jp^{2}$. There is spontaneous symmetry
breaking of of time-reversal symmetry, just as for a ferromagnet. The action (\ref{class}) is first-order in 
time derivatives. Thus, at least in the semiclassical approximation, the spin waves are not
Yang-Mills gluons, but nonrelativistic gauge Bosons. If the term $K\sum \gamma^{5}$ is included, this action
is no longer sufficient to describe spin waves. The Heisenberg equations of motion still indicate a nonrelativistic
relation between energy and momentum, however.  

There is a gauge magnet with relativistic gluons \cite{OR}. In $2+1$ dimensions, it has the form
\begin{eqnarray}
H=J_{1}\sum_{x^{1}+x^{2}\; {\rm even}}{\rm Tr}\;UUUU+J_{2}\sum_{x^{1}+x^{2}\; {\rm odd}}{\rm Tr}\;U^{5}U^{5}U^{5}U^{5},
\label{rel}
\end{eqnarray}
where each term is on an elementary plaquette. This is a staggered model on a chessboard, with one type of term on
red plaquettes, the other on black plaquettes. There is a similar model version in higher dimensions too.
The spin waves are similar to those of a one-dimensional spin chain \cite{ACS}, and
have speed of light 
$c=8{\sqrt{\vert J_{1}J_{2}\vert}}$
and mass gap
$m=\frac{\vert J_{1}-J_{2}\vert }{8\vert J_{1}J_{2}\vert}$, respectively. This model has been speculated
to be a Yang-Mills theory with a Chern-Simons term in $2+1$ dimensions \cite{OR}.

\section{Conclusions}

I've tried to make two points. The first is that gauge magnets may be ubiquitous. It
is fun to speculate that gauge invariance in particle physics arises this way, but I think not (a symmetry principle, like supersymmetry, is needed
to give all the gauge bosons the same speed of light). The 
second point is that better methods are needed for identifying the quantum
field theory described by a gauge magnet.

Coherent state methods \cite{OR} need to be generalized. The formalism used thus far does not properly acommodate
all the observables. Perhaps a Holstein-Primakoff method, analogous to that used for SU($2$) ferromagnets and 
antiferromagnets exists. Such a method would go a long way towards yielding a convincing correspondence with
a field theory. Such methods, of course, are only reliable only in the large-spin limit (in the models presented here, the spin is one-half). In the short term, perhaps this limitation is not so important. The ultimate development of such methods would be
a correspondence between gauge magnets and gauge field theories similar to Haldane's correspondence between spin chains
and the O($3$) sigma model \cite{Hal}.

{\bf Acknowledgements:} 
I would like to mention that Daniel Rohrlich first noticed there are also
relativistic spin waves in (\ref{hamil1}),
which are the photons of the U($1$) gauge symmetry (though we did not think about this symmetry, at that time). I also am grateful
for Alessio Celi, Luca Tagliacozzo and Maciej Lewenstein for teaching me about many of the advances which have occured
recently in this subject. I thank Benni Reznik and Yanick Meurice for useful comments. Finally, I thank Axel Cort\'{e}s Cubero
for recent discussions about the relativistic model (\ref{rel}).

\end{document}